\documentclass[aps,prl,reprint,superscriptaddress]{revtex4-1}
\usepackage{graphicx}

\begin{document}


\title{Avoiding Tokamak disruptions by applying static magnetic fields that\\
align locked modes with stabilizing wave-driven currents}


\author{F.A. Volpe}
\email[]{fvolpe@columbia.edu}
\affiliation{Dept Applied Physics and Applied Mathematics, 
Columbia University, New York, NY 10027, USA}

\author{A.~Hyatt}
\author{R.J. La Haye}
\author{M.J. Lanctot}
\author{J. Lohr}
\author{R. Prater}
\author{E.J. Strait}
\author{A. Welander}
\affiliation{General Atomics, P.O. Box 85608, San Diego, California USA}


\date{\today}

\begin{abstract}
Non-rotating (`locked') magnetic islands often lead to complete
losses of confinement in tokamak plasmas, called major
disruptions. Here locked islands were suppressed for the first time, 
by a combination of applied three-dimensional magnetic
fields and injected millimetre waves. The applied fields were used to
control the phase of locking and so align the island O-point with the
region where the injected waves generated non-inductive currents. This
resulted in stabilization of the locked island, disruption avoidance, 
recovery of high confinement and high pressure, in accordance with the
expected dependencies upon wave power and relative phase between
O-point and driven current.
\end{abstract}


\maketitle


The international ITER \cite{c1} tokamak has the objective of
demonstrating the scientific feasibility of magnetic confinement
fusion as a source of energy. A concern towards the achievement of
this goal is represented by major disruptions \cite{c1}: complete losses of
confinement often initiated \cite{c3} by a non-rotating (`locked') magnetic
island created by magnetic reconnection \cite{c4}. During disruptions,
energy and particles accumulated in the plasma volume over several 
confinement times (seconds in ITER, a fraction of a second in present 
experiments) are lost in a few milliseconds and released on the
plasma-facing materials \cite{c5}. In addition, multi-MA level currents
flowing in the tokamak plasma for its sustainment and confinement are
lost, also in milliseconds, thus terminating the plasma discharge and
causing electromagnetic stresses that, if unmitigated, could lead to
excessive device wear. Here it is shown for the first time 
that magnetic perturbations
can be used to avoid disruptions by ``guiding'' the magnetic island to
lock in a position where it is accessible to millimetre wave beams  
that fully stabilize it.  
Stabilization is due to locally wave-driven currents 
(Electron Cyclotron Current Drive, or ECCD).  

Magnetic control of island rotation \cite{DITE} and stabilization of rotating 
islands by 
ECCD \cite{c9} were separately demonstrated in the past. 
Currents were either 
continuously driven \cite{c9} or, more efficiently, they were modulated in 
synch with the spontaneous island rotation \cite{c10}. 
Electron Cyclotron Heating (ECH) was also used for stabilization 
\cite{Classen}, but is predicted to scale unfavorably to large hot plasmas 
\cite{c18}. 
Two experiments combined magnetic perturbations -to {\em produce} the island-  
with ECH that stabilized it: in the first one the mode was born locked to 
a given phase and was  
stabilized by continuous ECH \cite{Morris2}; the second one controlled island 
rotation and stabilized the mode by modulated ECH \cite{TEXTOR}.

However, if the rotating island (typically a spontaneous, pressure-driven 
'Neoclassical Tearing Mode') 
is not preempted or stabilized (due for example to late
intervention, misalignment, or insufficient power being used for this
purpose), or if the island does not ever rotate at all, it becomes
necessary to suppress the locked mode. This capability was 
numerically modeled \cite{Yu}, experimentally tested \cite{c20}, and is 
fully demonstrated here for the first time. 
Without this capability, the locked mode would grow and promptly
lead to a disruption. After that, only one last line of defense would
remain, namely to mitigate the disruption, for instance by massive gas
injection \cite{c1, c11}.  

{\em Static}, rather than rotating fields \cite{TEXTOR,c20}, are used here, 
permitting to align the plasma such that ECCD can be 
continuously deposited into the location of the locked mode where it has a 
stabilizing effect on it.  
By contrast, continuous ECCD on rotating islands is always on, but  
only stabilizing half of the time \cite{c9,c20}, 
and modulated ECH/ECCD on rotating islands is on and stabilizing half of the 
time, if properly phased \cite{c10,TEXTOR}. 

Locked mode control will be needed in ITER, where
(1) islands will be locked for most of their lifetime and (2)
alignment will be challenging. This is because islands of poloidal/toroidal
mode number $m/n$=1 are expected to lock as soon as they exceed a
width of 5 cm, i.e. seconds after forming \cite{c12, c13} and well before
reaching an ultimate width of 35-40 cm \cite{c14}. Hence, it will be
challenging to precisely aim the ECCD for preemption or in the
brief period of time when the island is still rotating, but very
small. This will require few cm of precision at several meters from
the wave launcher, resulting in 0.2 degrees of angular precision \cite{c15},
whereas locked modes are larger, easier targets. In addition, 
rapid locking sets a requirement for
rapid mirror steering, if one wants to align the ECCD and stabilize the mode 
when still rotating.

The locked island O-point (i.e. the local magnetic axis of
the island) can lock in a
position not necessarily accessible to the mm-waves. In the
absence of position control, one can apply ECH 
only (no current drive) to delay or avoid disruptions in small \cite{c16}
and mid-size devices \cite{c17}. This approach, as mentioned, is predicted to
scale unfavourably to larger, hotter fusion plasmas such as ITER,
where the stabilization is expected to be completely governed by
current drive \cite{c18}. Yet, in order for current drive to be used, it is
necessary to gain control of the locking position of the magnetic
island, as currents driven at the wrong location, such as the island
X-point (the tip of the island), can actually be destabilizing \cite{c19}.

The island is caused by a helical ``hole'' in a pressure-driven
(``bootstrap'') current \cite{c18} and 
can be modelled with multiple
helical filaments offset in the toroidal angle and carrying different
currents, according to a sinusoidal distribution \cite{Shiraki}. 
A magnetic dipole is associated with this helical current pattern.  
When not controlled, the magnetic dipole of an 
initially rotating $n$=1 mode, slowed down by the interaction with the
currents induced in the resistive wall of the tokamak,
tends to align with the $n$=1 ``error'' in the otherwise axisymmetric 
tokamak field. Error fields as small
as one part in $10^4$-$10^5$ of the main toroidal field are sufficient to
cause a pre-existing rotating island to lock, or to directly cause a
non-rotating (locked) island to form.  

In both cases, the azimuthal angle (toroidal phase) of the locked island is 
determined by the error field. 
In previous work, slowly rotating perturbations 
were superimposed to it, and the resultant acted as ``magnetic tweezers'' that
slowly rotated the locked mode \cite{c20}. In the present work, {\em static}  
$n$=1 magnetic perturbations are applied as soon as a rotating magnetic
island is magnetically detected to decelerate, before it comes to a
complete stop. 
Their amplitude and toroidal phase are chosen in
such a way that, when the island locks to the total $n$=1 field, 
its O-point is toroidally aligned with the ECCD deposition region. 
The optimal
perturbation yielding good alignment can either be calculated in
advance (if the error field is known), be experimentally optimized (as
in the experiments presented here), or it can be chosen to be strong
enough as to dominate over the error field in determining the toroidal
phase of the island, $\phi$. In any case, it should be mentioned that, due
to the low toroidal number of the mode, $n$=1, the optimal ``target''
for the mm-waves is toroidally elongated. Therefore, once a means of
controlling $\phi$ is available, the necessary precision in $\phi$ is relatively
low, of the order of $\pm 45^o$. With this {\em phase} control in hand, the
current drive mechanism, which can be stabilizing or destabilizing,
can be effectively used to control the locked mode {\em amplitude} as
desired, i.e. to fully stabilize the mode and avoid the disruption. 

\begin{figure}[!htb]
  \includegraphics{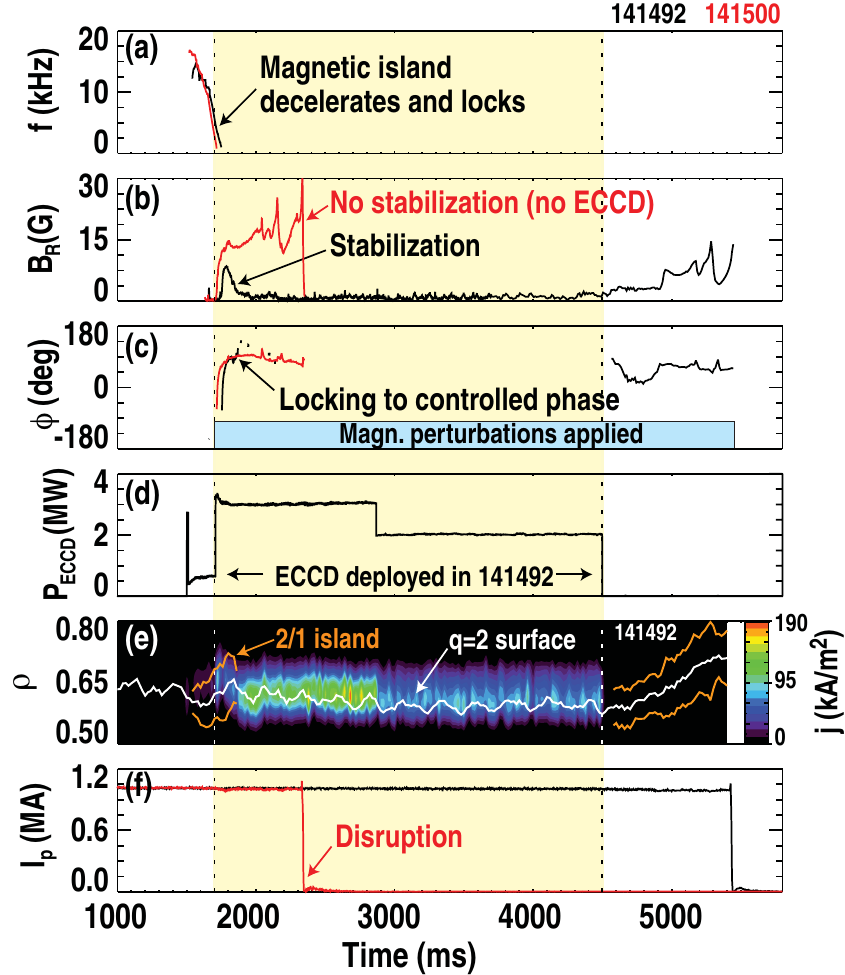}
  \caption{Locked Mode Stabilization. A rotating $n$=1 Neoclassical Tearing 
    Mode island is
    detected to slow down by magnetic probes (a). Shortly thereafter a
    locked island is detected by saddle coil sensors of the radial
    magnetic field $B_R$ (b). A control system reacts to deceleration 
    by applying a static $n$=1 field. The resultant of this
    perturbation and of the machine ``error field'' causes the island to
    lock with a phase (c) such that ECCD power (d), also injected in response 
    to island deceleration, generates non-inductive currents
    (e, colour contours) in the island O-point, 
    i.e. in the correct toroidal location, as well as in the correct
    radial location normalized to the plasma minor radius, $\rho$ (e,
    white). The island edges displayed (e, orange) are based on the
    calculated island width. The calculation \cite{c29} is based on the
    measured poloidal and radial field, respectively, when the island
    is rotating or locked. The result is
    stabilization of the locked island (b, black) and disruption avoidance, as 
    indicated by the plasma current (f). In an 
    otherwise identical discharge, but without ECCD, 
    the locked island is not stabilized and causes the plasma to disrupt (red). 
    \label{fig2}}
\end{figure}

Fig.\ref{fig2} shows two plasma discharges realized
at the DIII-D tokamak \cite{c22} and characterized by locked modes. In the
discharge depicted in black, the simultaneous use of magnetic
perturbations for phase-control and of mm-waves for amplitude-control
resulted in rapid stabilization of the locked mode (Fig.\ref{fig2}b). The
stabilization is considered complete because the radial field signal
measured with inductive sensors \cite{c23} decreases to the 1 G level, which
is consistent with noise and with other $n$=1 activity. Importantly, the
disruption is avoided for as long as both controls are deployed. If
magnetic perturbations alone are used (discharge depicted in red) or
mm-wave current drive is turned off (black discharge at time $t$=4500
ms), the mode grows and disrupts the discharge, as exemplified by the
dramatic drop in plasma current (Fig.\ref{fig2}f).

\begin{figure*}[!htb]
  \includegraphics{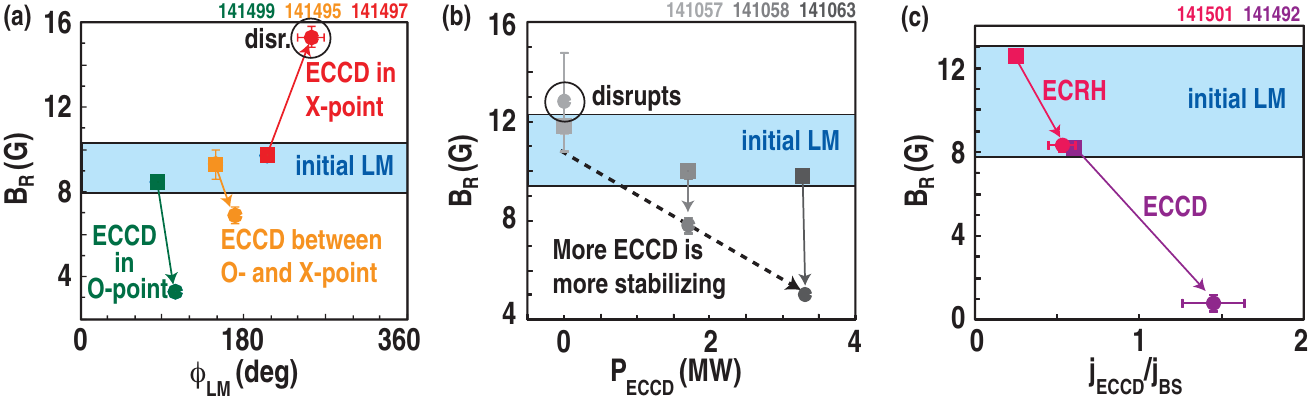}
  \caption{Evolution of locked mode amplitude -as measured at the wall, 
    in the radial component- as function of (a) the toroidal location of 
    locking $\phi_{LM}$, (b) the applied mm-wave power $P_{ECCD}$, and (c) the 
    density of EC-driven current, $j_{ECCD}$, normalized to the local 
    ``bootstrap'' current density $j_{BS}$. Such ratio is higher for  
    ECCD-optimized than for ECH-optimized wave launch; 
    deposition was in the O-point for both cases. Squares in the blue stripe 
    denote the amplitudes of the initial locked modes (LMs).  
    Filled circles denote the ``final'' 
    amplitudes, after stabilization -if any- or prior to disruption. 
    Arrows pointing downwards or upwards indicate respectively stabilization 
    or destabilization. Cases resulting in disruptions are circled and labeled. 
    \label{fig3}}
\end{figure*}

It is also experimentally confirmed that: 
\begin{enumerate}
\item
To be stabilizing, the driven
current needs to be deposited in the O-point of the locked island,
whereas deposition in the X-point is destabilizing. The effect of
deposition in an intermediate location is, indeed, intermediate,
i.e. neither strongly stabilizing nor strongly destabilizing
(Fig.\ref{fig3}a). Note that the three cases plotted differ by the
orientation of the applied non-axisymmetric field, and thus by the
amplitude of the total non-axisymmetric field. However, this was
recently found to have negligible effect on the locked island
\cite{c24}. Also note that the island moves slightly after locking. Its
initial phase is magnetically controlled by the applied fields, which
are subsequently kept constant. The change of phase is ascribed to
changes in the error field and in the viscous, neutral beam and
electromagnetic torques acting on the island, reaching balance at a
new phase. In turn, such torques change as a consequence of the very
fact that the island size and island current evolve.  

\item
Higher
mm-wave power, all the rest remaining the same, has a more stabilizing
effect (Fig.\ref{fig3}b). This is due to more intense stabilizing currents
being driven.  

\item 
The key requisite for stabilization is that the
applied wave-driven current compensates or overcompensates for the
missing pressure-driven bootstrap current responsible for the
formation of the island. To this end, it is more efficient to use the
available wave power to drive maximum current. This is obtained for
injection at an appropriate oblique angle relative to the magnetic
field \cite{c25}. If, instead, injection is perpendicular or nearly
perpendicular to the field, the main effect is some heating but little
or no current, which typically is insufficient for complete
stabilization (Fig.\ref{fig3}c).  
\end{enumerate}

In addition to disruption avoidance,
locked mode control provides benefits for confinement, compared with
non-stabilized discharges. At the same time, an increase in
confinement represents an additional, indirect evidence that the
island was stabilized. The reason is that a large island alters the
magnetic topology in a way that creates a local short circuit
for heat and particles, degrading confinement \cite{c26}. Therefore, the
suppression of the large locked island restores good particle and
energy confinement, density $n$, temperature $T$ and their product (the
kinetic pressure), as well as high normalized pressure $\beta _N$, 
defined as the ratio between the kinetic
pressure of the plasma and the magnetic pressure used to confine it,
normalized to $I_p/aB$, where $I_p$ is the plasma current in mega-amperes, $a$ 
the minor radius in meters and $B$ the magnetic field in Tesla.  

\begin{figure}[!b]
  \includegraphics{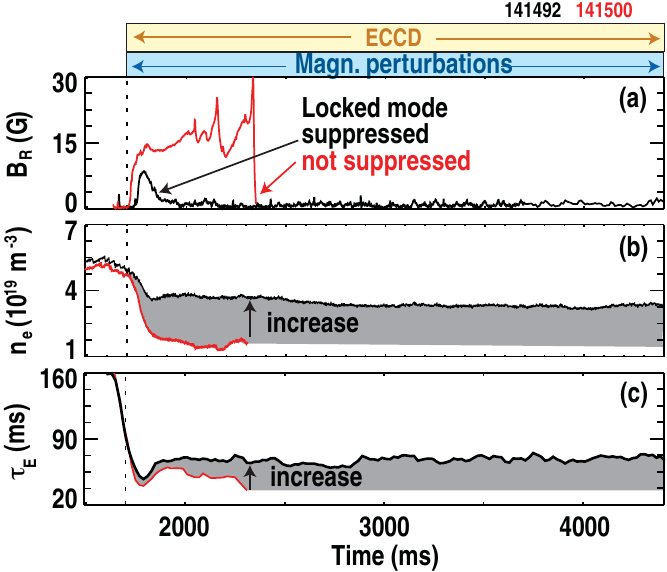}
  \caption{Increase of confinement. Suppressing the $n$=1
    locked mode (a, black) improves particle and energy confinement
    over the unsuppressed case (red), as evident for example from the
    electron density (b) and energy confinement time (c).
    \label{fig4}}
\end{figure}

\begin{figure*}[!htb]
  \includegraphics{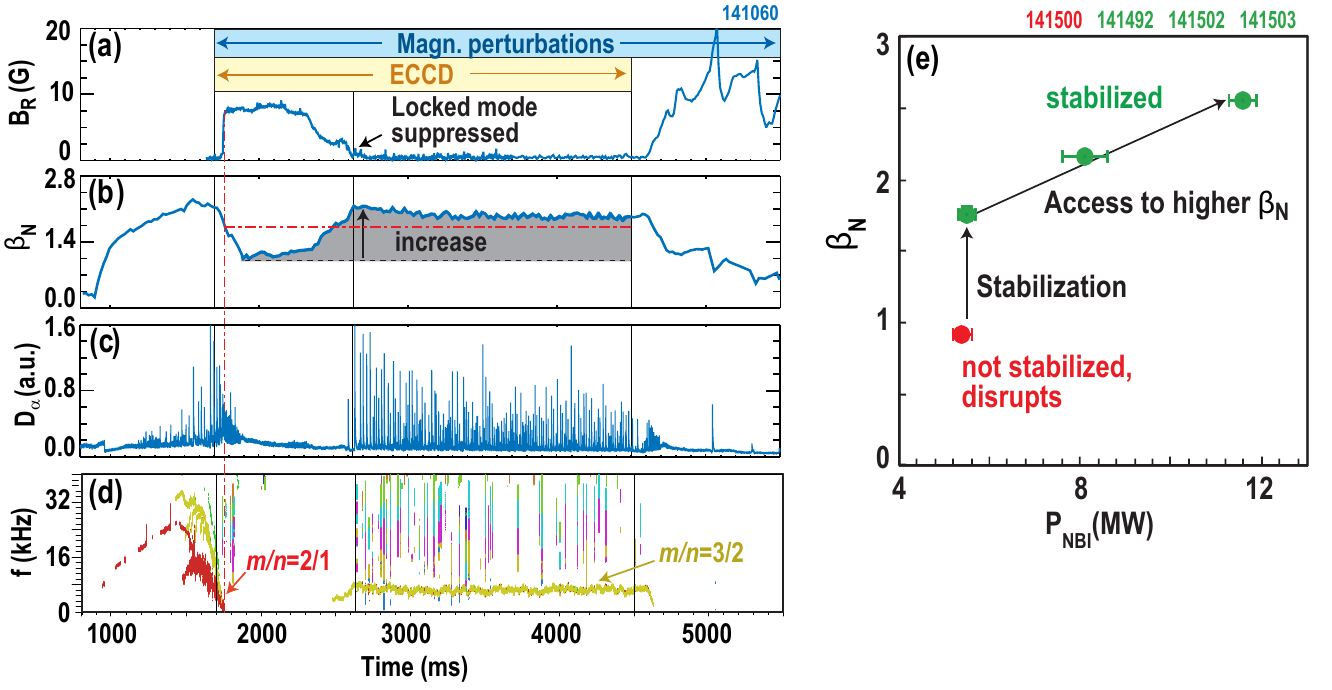}
  \caption{a-d.  
    Red dash-dotted lines mark the time of $n$=1 locking and the
    value of $\beta_N$ at that time. After stabilization, $\beta _N$ returns to
    values comparable with the original one, and is only limited by
    the onset of a pressure-driven, pressure-limiting but
    non-disruptive rotating island of poloidal/toroidal numbers
    $m/n$=3/2. Such island is visible in the spectrogram of mode
    amplitude as a function of time and frequency (d, yellow). 
    Bursts of broadband activity in panel d are due to Edge Localized Modes, 
    also recognizable as bursts of Balmer-alpha emission (c), and indicative 
    of high confinement (H) mode. Note the H-mode being re-established after 
    locked mode stabilization.  e. The
    effect of stabilization at fixed Neutral Beam
    Injection (NBI) heating power is a marked increase in $\beta _N$. If,
    further to that, the NBI power is increased, $\beta _N$ grows accordingly,
    as expected. These high values of power and $\beta _N$ were not accessible
    in non-controlled discharges, which were limited by disruptions to
    $\beta_N \approx$ 0.9.
    \label{fig5}}
\end{figure*}

The suppression of the locked island allows an edge transport barrier to
re-form, thus re-establishing the so called high confinement mode, or
H-mode \cite{c27} (Fig.\ref{fig5}c). As a result, higher electron density 
$n_e$ and energy confinement time $\tau_E$ are achieved when the locked
mode is suppressed (Fig.\ref{fig4}, black), compared with when the mode is not
suppressed (red). Two factors, though, limit the increase of $\tau_E$. One
is that the mm-wave power is intended for mode suppression, not
heating.  It is deposited near the plasma edge and quickly lost. The
other is that $n$=1 magnetic perturbations are still applied after
locked mode suppression and, similar to error fields, they have an
impact on confinement. Furthermore, it is not the case in Fig.\ref{fig4}, 
but other modes appearing after suppression of the $n$=1 mode can also limit
confinement.  

Fig.\ref{fig5} documents the effect of locked islands and
their stabilization on $\beta_N$. A rotating $n$=1 island appears at 1460 ms
(Fig.\ref{fig5}d). As a result, and in spite of the Neutral Beam Injection
(NBI) heating power being increased (not shown), $\beta_N$ decreases, and 
keeps decreasing after the island locks (Fig.\ref{fig5}b). 
ECCD, however, stabilizes the island (Fig.\ref{fig5}a) 
and leads to high values of
normalized pressure: for the same amount of NBI (5 MW), $\beta_N$ drops as
low as 1.1 after locking and grows as high as 2.2 after stabilization
(Fig.\ref{fig5}b).
As mentioned, locked mode stabilization also re-establishes the H-mode, 
as indicated by the presence of Edge Localized Modes in 
Balmer-alpha emission ($D_\alpha$, 
Fig.\ref{fig5}c). The H-mode is maintained for as long as the ECCD is 
deployed, and is lost after the ECCD is turned off and the mode reappears 
(Fig.\ref{fig5}a,c).

Note that the locked $n$=1 island is really suppressed, not unlocked: after
the loss of locked mode signal in Fig.\ref{fig5}a, no rotating 2/1 island
reappears in Fig.\ref{fig5}d for as long as ECCD is deployed
at the island location. The very fact that the pressure becomes
high again, however, may lead to other pressure-driven,
pressure-limiting instabilities appearing elsewhere in the plasma. A
common example is the rotating 3/2 island in Fig.\ref{fig5}d. Without that, 
$\beta_N$
might have reached even higher values, or at a lower ``cost'' in
terms of NBI power (2.4 MW earlier in the same discharge, before any
island had appeared). The stabilization of the rotating 3/2 island is
well established \cite{c9,c19}, and goes beyond the scope of the present
work.  

Finally, Fig.\ref{fig5}e indicates that, for sufficiently high NBI
heating power, locked-mode-controlled discharges attain values of $\beta_N$ 
as high as 2.6 without terminating in disruptions. Equivalent
discharges with uncontrolled locked modes disrupt at low NBI power and
$\beta_N$ as low as 0.9. 
Note that values of $\beta_N \ge$2.6 were obtained in the past in discharges not
subject to locked modes \cite{c28}.

In summary, applied non-axisymmetric magnetic perturbations were used to
control the phase of locking of an initially rotating magnetic island. 
This permitted Electron Cyclotron Current Drive stabilization of the locked
island, which avoided the plasma 
disruption and re-established the high confinement (H) mode. 

It is important to note that the technique makes use of static magnetic
perturbations that need to penetrate in the plasma on a relatively
benign timescale. The estimated locked mode growth-rate in ITER
(1.1cm/s, with saturation at up to 35-40 cm) \cite{c13} and slowdown time
before locking (4 s) give ample time for an externally applied static
field to penetrate through walls that, in ITER, will have an $n$=1
resistive time of 190 ms. As a consequence it should be possible to
apply the desired $n$=1 static perturbation by means of
error-field-correction coils external to the vessel. This permits
dedicating the internal coils to tasks needing proximity to the plasma
and/or fast response, such as controlling edge localized modes and
vertical instabilities.

This material is based upon work supported by the U.S. Department of
Energy, Office of Science, Office of Fusion Energy Sciences, using the
DIII-D National Fusion Facility, a DOE Office of Science user
facility, under Awards DE-SC0008520 and DE-FC02-04ER54698. DIII-D data
shown in this paper can be obtained in digital format by following the
links at https://fusion.gat.com/global/D3D\_DMP. The help of 
H. Reimerdes in programming the neutral beam injection
is gratefully acknowledged, as are fruitful discussions with
R. Buttery, R. Groebner and J. Hanson.

\subsection{}
\subsubsection{}

\end{document}